\documentclass[apjl]{emulateapj}
\usepackage{mathrsfs}
\usepackage{amssymb}
\usepackage[usenames,dvips]{color}
\usepackage{ulem}

\newcommand{\od}[2]{\frac{d #1}{d #2}}
\newcommand{\pd}[2]{\frac{\partial #1}{\partial #2}}
\newcommand{\prn}[1]{\left ( #1 \right )}
\newcommand{\brk}[1]{\left [ #1 \right ]}
\newcommand{\Del}{\mathbf{\nabla}}
\def\msun{{\rm ~M}_{\odot}}
\begin{document}

\title{The Equilibrium Structure of Prolate Magnetized Molecular Cores}
\author{Michael J. Cai\altaffilmark{1} and Ronald E. Taam\altaffilmark{1,2,3}}
\altaffiltext{1}{Academia Sinica Institute of Astronomy and Astrophysics-TIARA, P.O.
Box 23-141, Taipei, 10617 Taiwan}
\altaffiltext{2}{Academia Sinica Institute of Astronomy and Astroophysics/National
Tsing Hua University-TIARA, Hsinchu Taiwan}
\altaffiltext{3}{Northwestern University, Department of Physics and Astronomy,
2131 Tech Drive, Evanston, IL 60208}

\begin{abstract}
The structure of molecular cloud cores supported by thermal pressure and a poloidal magnetic field
is reinvestigated in the magnetohydrostatic and axisymmetric approximation. In addition to oblate
configurations found in earlier work, solutions yielding prolate spheroidal shapes have also been
obtained for a reference state described by a uniform sphere threaded by a uniform background magnetic
field. The solutions for prolate configurations are found to be relevant for lower masses than for
their oblate counterparts. Of particular importance is the result that the prolate cloud cores have
radii less than a maximum given by $0.25 pc \prn{\frac{a}{0.2km/s}}^2 \prn{\frac{P_{\rm ext}}{10^{-12}
dyne/cm^2}}^{-1/2}$, where $a$ is the sound speed and $P_{\rm ext}$ is the external pressure of the
background medium. The existence of such solutions obviates the presence of toroidal fields in such
modeled structures.
\end{abstract}
\keywords{star formation, molecular cloud cores, magnetic fields}

\section{Introduction}
The study of dense molecular cloud cores are of paramount interest because they are the sites where
stars form \citep[see][for reviews]{Shu:1987, McKee:2007}. In situations where other physical effects
in comparison with gravity and thermal pressure can be neglected, these cores have been modeled as
isothermal spheres in hydrostatic equilibrium bounded by an external pressure, known as Bonnor-Ebert
spheres \citep{Bonnor:1956, Ebert:1955}. For some objects, such a model offers remarkable agreement
with observation \citep{Alves:2001, Evans:2001, Kirk:2005, Schnee:2005, Stutz:2007}. However, surveys of large samples of dense cores in dark clouds reveal that spherically symmetric cores are the exceptions rather than the rule. The projected aspect
ratios can significantly differ from unity, making the intrinsic geometry even more elongated
\citep{Myers:1991,Jijina:1999}. If we assume these cores are axisymmetric and are randomly oriented in the sky,
statistical analysis suggests many of them are prolate \citep{Myers:1991, Ryden:1996}. On the other
hand, oblate shapes may fit better with the observed distribution of shapes for intrinsically triaxial
cores \citep{Jones:2001, Tassis:2007}.

The formation of prolate cores presents certain theoretical challenges. If the formation history of
these cores is dominated by quasi-static contraction regulated by ambipolar diffusion, the resulting
geometry is a sequence of oblate spheroids \citep{Mouschovias:1976a, Mouschovias:1976b, Nakano:1979,
Lizano:1989}. This is a natural outcome since matter can contract freely along the field lines, but
must overcome the additional magnetic pressure and tension perpendicular to them. The inclusion of
rotation produces even flatter morphologies, but there is little observational evidence supporting the
view that rotation is important for cores \citep{Goodman:1993}. \citet{Tomisaka:1991} and
\citet{Fiege:2000} invoked a helical magnetic field to explain the prolate cores and were able to
produce a large range of aspect ratios consistent with observation.  Here, the hoop stress generated
by the toroidal field component facilitates the confinement of matter toward the axis.  However, the
presence of toroidal fields requires a current flowing along the symmetry axis, which is difficult
for astrophysical systems to realize. We note that \citet{Curry:2001} were able to obtain both prolate
and oblate cores with only a poloidal field by specifying the shape of the core and solving for the
mass-to-flux ratio after an equilibrium had been found. In this case, the mass-to-flux ratios are
distinct for the two types of cores.

In this Letter, we reexamine the structure of molecular cloud cores, which are treated as
magneto-hydrostatic equilibria, with the ultimate goal of providing a grid of models that can be used
for interpreting pre-stellar core density structures.  We have found that magnetized structures for
both prolate and oblate configurations can be produced for a wide range of aspect ratios from a purely
poloidal field configuration with a generic functional form of the mass-to-flux ratio. This is in
contrast to previous work where prolate configurations were produced only with a toroidal field or a
specific mass-to-flux ratio.  The findings of this work help to elucidate the physical parameters that
dictate the morphologies of dense molecular cores.  The theoretical framework including the assumptions
and equations of the model are described in \S 2. The method of solution and the numerical results are
presented in \S 3. Finally, we summarize and discuss the results in \S 4.

\section{Formulation and Basic Equations}
For mathematical tractability, the figures of equilibrium are restricted to be axisymmetric and
described in cylindrical coordinates ($\varpi$, z).  Consider
a cloud core described by an isothermal equation of state $P = a^2 \rho$, where $P$, $\rho$, and $a$
are the pressure, density, and a constant sound speed. Let this cloud be embedded in an environment
characterized by a large scale background magnetic field $\mathbf{B}_0 = B_0 \mathbf{\hat z}$ and an
external pressure $P_{\rm ext}$. Pressure equilibrium requires the cloud to have density $\rho_0 =
P_{\rm ext}/a^2$ on its surface. From these parameters, along with the gravitational constant, $G$, the
units of length, density, pressure, gravitational potential, mass, magnetic field, and magnetic flux
are taken as
\begin{eqnarray*}
  &l_0 \equiv \frac{a}{\sqrt{4\pi G \rho_0}}, \qquad \rho_0, \qquad P_{\rm ext}, \qquad a^2, \\
  &M_0 \equiv 4\pi \rho_0 l_0^3, \qquad B_0, \qquad \Phi_0 \equiv 2\pi B_0 l_0^2,
\end{eqnarray*}
respectively.  For typical cores, if we take $a = 0.2$ km s$^{-1}$, $P_{\rm ext} = 10^{-12}$ dyne
cm$^{-2}$, and $B_0 = 20\mu G$, then $l_0 = 0.14 pc$, $\rho_0 = 1.5\times 10^3 m_H$ cm$^{-3}$,
$M_0 = 1.3\msun$, and $\Phi_0 = 2.5 \mu$G pc$^2$, where $m_H$ is the mass of a hydrogen atom.  In
these units, the magneto-hydrostatic equilibria are governed by Poisson's equation for the
gravitational potential, V, as
\begin{equation}
  \nabla^2 V = \rho, \label{Poisson}
\end{equation}
the divergence free condition of the magnetic field
\begin{equation}
  \Del \cdot \mathbf{B} = 0,
\end{equation}
and the force equation
\begin{equation}
  0 = -\Del \log \rho - \Del V + (\beta \rho)^{-1}(\Del \times \mathbf{B}) \times \mathbf{B}.
\end{equation}
In the last equation, $\beta \equiv 4\pi \rho_0 a^2/B_0^2$.  The other two equations of magnetohydrodynamics
(MHD), the conservation equation, and the induction equation are identically satisfied for $\partial_t =
\mathbf{v} =0$.

The divergence free condition on the magnetic field can be satisfied using the flux function $\Phi = \int
\mathbf{B} \cdot \mathbf{e}_z \varpi d\varpi$ (recall a factor of $2\pi$ has been absorbed into the unit of
flux).  An axisymmetric and purely poloidal magnetic field can be uniquely specified by the flux function as
\begin{equation}
  \mathbf{B} = \varpi^{-1} \Del \Phi \times \mathbf{e}_\varphi.
\end{equation}
In this case, $\mathbf{B}$ is everywhere tangent to the contours of $\Phi$.  In terms of the flux function,
the force equation becomes
\begin{equation}
  0 = -\Del \log \rho - \Del V - (\beta \rho)^{-1} \Del \cdot (\varpi^{-2} \Del \Phi) \Del \Phi.
\end{equation}
The force equation can be projected along the magnetic field to obtain $\mathbf{B} \cdot \Del h = 0$, where
\begin{equation}
  h \equiv \log \rho + V, \label{enthalpy}
\end{equation}
is the specific enthalpy. This implies that $h(\varpi, z) = h(\Phi)$ is a function of $\Phi$ alone.  The
component of the force equation perpendicular to the magnetic field, known as the Grad-Shafranov equation,
governs the spatial distribution of field lines, and it can now be manipulated to read
\begin{equation}
  \Del \cdot (\varpi^{-2} \Del \Phi) = -\beta \rho \frac{d h}{d\Phi}.\label{GSE}
\end{equation}
To close the system of equations, we impose the integral constraints that the mass in each flux tube is conserved, and is given by
\begin{equation}
  \int_0^{Z(\Phi)} \rho \varpi \pd{\varpi}{\Phi}\Big \vert _z dz = \frac{dm}{d\Phi}.
\end{equation}
In the above equation, the $z$ integral is performed over constant $\Phi$, and $dm/d\Phi$ is the known differential mass-to-flux ratio, obtainable from either observation or an evolutionary calculation.  The core
surface, described by $Z(\Phi)$, is a free internal surface of the problem.  To determine the location
of the core boundary and the specific enthalpy, we note that $h = V \Big \vert_{z=Z}$ because $\rho = 1$
on the surface.  Once $V$ is known, both $h$ and $Z$ can be determined by solving the equation
\begin{equation}
  e^V\Big \vert_{z = Z} = \od{m}{\Phi} \brk{\int_0^{Z(\Phi)} e^{-V} \varpi \pd{\varpi}{\Phi} dz}^{-1},\label{compute_h}
\end{equation}
along each flux tube.

\section{Numerical Methods and Results}
Our numerical scheme for constructing solutions is an iterative procedure similar to that of
\citet{Mouschovias:1976a} and \citet{Tomisaka:1988a}. For definiteness, we adopt a differential
mass-to-flux ratio corresponding to a reference state consisting of a uniform sphere threaded by the
uniform background field.  If the total mass of the core is $M_c$, and total trapped flux is $\Phi_c$, then
\begin{equation}
  \od{m}{\Phi} = \frac{3M_c}{2\Phi_c} \sqrt{1 - \frac{\Phi}{\Phi_c}}. \label{init_state}
\end{equation}
This configuration was termed ``parent cloud'' by the authors, and serves as the initial trial solution.
Starting with the parent cloud, $h$ and $Z(\Phi)$ are computed from equation (\ref{compute_h}), which
allows a calculation of the source functions in equations (\ref{Poisson}) and (\ref{GSE}). These equations
are then solved by successive over relaxation to obtain intermediate solutions.  A corrected solution is
constructed by under relaxation and updated with the next iteration. The procedure continues until the
fractional error is reduced to a pre-specified level. Once a converged solution is obtained, higher accuracies are then achieved by a series of mesh refinements.

We were able to reproduce the solutions presented by \citet{Mouschovias:1976b} and \citet{Tomisaka:1988b} using their input parameters.  We compare our result visually to theirs and estimate the fractional error to be within 5\%.  In particular, for a fixed total trapped flux, there is a maximum mass beyond which no equilibrium solution exists.  The numerical value of this maximum mass agrees with earlier work by \citet{Tomisaka:1988b} to within 3\%. This mass is equivalent
to the Bonnor-Ebert mass, the maximum stable mass for a given temperature and external pressure, but
modified by the presence of a magnetic field.

In the magnetically regulated star formation paradigm, a key parameter is the total mass-to-flux ratio,
$\lambda = \beta^{1/2}M_c/\Phi_c$ (or in conventional units $\lambda = 2\pi G^{1/2} M_c/\Phi_c$).  The
supercritical clouds (with $\lambda > 1$) are capable of continued contraction leading to star formation,
while the subcritical clouds (with $\lambda < 1$) are not.  Through a process of natural selection,
modern day molecular clouds are most likely to be in a marginally critical state \citep{Shu:2004}.  Clouds
with $\lambda \gg 1$ would have collapsed to form stars, while those with $\lambda \ll 1$ would have evolved
to the diffuse interstellar medium.  These theoretical arguments are consistent with the findings of \cite{Troland:2008} who estimated $\lambda \sim 2$ in the cores.  Thus, we argue that instead of fixing the total flux, as in \citet{Mouschovias:1976b} and \citet{Tomisaka:1988b}, a more convenient parameterization would be to vary the total mass while maintaining an order unity value of $\lambda$.

The solution space is three dimensional, parameterized by $\lambda$, $\beta$, and $M_c$. However, for
values of $\lambda \in [0.1,7]$ and $M_c < 20$, the solution is approximately
degenerate. With $\alpha \equiv \lambda/\beta^{1/2} \equiv M_c/\Phi_c$ held fixed,
varying $\beta$ by two orders of magnitude only introduces less than a 1\% change to the equilibrium
solutions, indicating the relative insensitivity of the magnetic field strength in this parameterization.
We shall thus use $\alpha$ and $M_c$ as our basic parameters when discussing the numerical results.

\begin{figure}[ht]
\begin{center}
\includegraphics[angle=90, width = 3.5in]{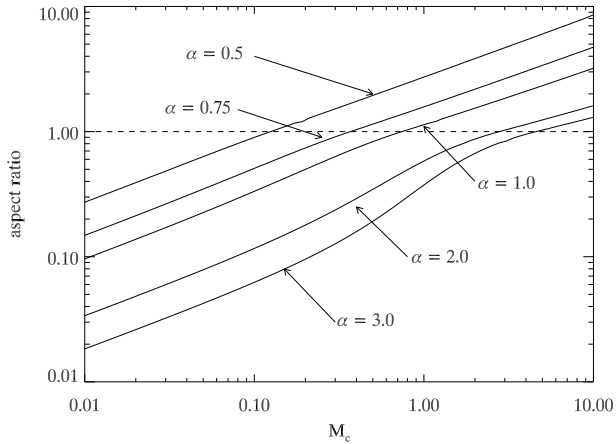}
\end{center}
\caption{Core aspect ratios as functions of cloud mass for $\lambda = 1$, and $\alpha = 0.5, 0.75,
1.0, 2.0, 3.0$.  The dashed line indicates the critical mass at which the aspect ratio is unity.}\label{aspect_fig}
\end{figure}

\begin{figure}[ht]
\begin{center}
\includegraphics[angle=90,width = 3.5in]{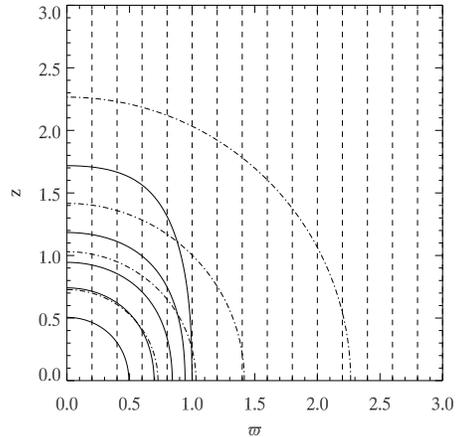}
\caption{A typical prolate core solution is illustrated in cylindrical coordinates ($\varpi$, z). The solid
curves are density contours, the dashed curves represent magnetic field lines, and the dash-dotted curves
are contours of the gravitational potential. For this particular solution, $\lambda = 1$, $\alpha = 2$,
and $M_c = 1$.  It has an aspect ratio of $0.582$, a central density of $\rho_c = 2.5$, and a radius of
$R = 1$.}\label{density_fig}
\end{center}
\end{figure}

Of primary interest in this study is the core shape, and the aspect ratios of the cores are illustrated in
Fig. \ref{aspect_fig} as functions of $M_c$ for several values of $\alpha$.  Here, the aspect ratio is
defined as the ratio of the radial extent to the vertical extent of the cloud core surface. An important
feature to note is that for fixed $\alpha$, there exists a critical mass $M_{\rm crit}$ below which the
core takes a prolate shape.
It can be seen that the prolate configurations are restricted to low core masses.  Specifically, for $\alpha = 1$ and typical core parameters (see above), the critical mass is $0.96 \msun$. Furthermore, as $\lambda$ increases for fixed $\beta$ or as $\beta$ decreases with fixed $\lambda$, the critical core mass increases.
The density and magnetic field structure for a typical prolate solution is displayed in Fig.
\ref{density_fig}.

The solutions for prolate cores are also found to exhibit a maximum radius, beyond which such solutions
do not exist. In particular, the parent cloud radius, $R_c$ and the mass $M_{\rm crit}$ that leads to a
final core of unit aspect ratio are shown as functions of $\alpha$ in Fig. \ref{crit_fig}. Here, the final
core radius is virtually indistinguishable from that of the parent cloud. It is evident that the radius
is a nonmonotonic function of $\alpha$, in contrast to the core mass, exhibiting a maximum at a radius
of 1.75 (or about 0.25 pc for typical core parameters) for $\alpha \sim 3$.

\section{Summary and Discussion}
We have reexamined the structure of dense molecular cloud cores supported, in part, by magnetic forces
and thermal pressure. In most of the parameter space that we have surveyed, our results agree with
previous studies that exist in the literature. Of particular interest is the discovery of solutions for
which the cores assume prolate shapes.  As we have shown (see Fig. \ref{aspect_fig}), for a fixed
mass-to-flux ratio, prolate configurations are more relevant for smaller masses, while the oblate ones
are for larger masses.

The formation of prolate cores can be understood in an evolutionary sense not too different from that
of oblate ones.  For our choice of the reference state, and a fixed $\alpha$, the core mass is related
to the density, $\rho_i$, by
\begin{equation}
  M_c = \prn{\frac{\alpha}{2}}^3 \prn{\frac{3}{\rho_i}}^2.
\end{equation}
Cores with sufficiently small mass represent over dense regions, and they must expand to achieve
equilibrium. Because of the additional external pressure exerted by the magnetic field, the gas
experiences less resistance in the direction parallel to the field lines than perpendicular to them. The
subsequent evolution naturally leads to a prolate figure. Instead of an hour glass shaped magnetic field
commonly seen accompanying oblate cores in either theoretically calculations \citep{Mouschovias:1976b,
Tomisaka:1988b} or observations \citep{Girart:2006, Girart:2009}, the magnetic field lines bow outwards
in the midplane.  However, for our particular choice of the parent cloud, because the overall magnetic energy is much larger than thermal energy for small masses, the outward bowing of the field lines is not pronounced in general.  Of all the prolate shaped models we have constructed, the fractional distortion of the magnetic field lines only occurs at a $10^{-4}$ level.  Other forms of $dm/d\Phi$ may lead to more drastic outward bowing of the field lines for prolate cores \citep[see e.g.,][]{Curry:2001}.

\begin{figure}[ht]
\begin{center}
\includegraphics[angle=90, width = 3.5in]{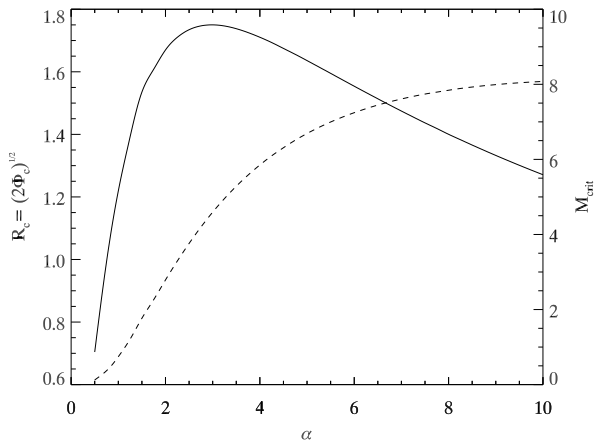}
\end{center}
\caption{Parent cloud radius (solid curve) and core mass (dashed curve) for solution with unit aspect ratios.}\label{crit_fig}
\end{figure}
It is instructive to inquire why previous studies, such as \citet{Mouschovias:1976b} and
\citet{Tomisaka:1988b}, did not find such kind of prolate cores with just poloidal fields, since they
have surveyed a large portion of the parameter space spanning many orders of magnitude in both $\beta$
and central density $\rho_c$.  The answer lies in the total flux trapped by the core.
\citet{Tomisaka:1988b} used a uniform sphere with various mass-to-flux distributions as a parent cloud
to initiate their iterative calculation.  Our particular choice of parent cloud is threaded by a uniform
field, which corresponds to their $N=1$ model. Its radius is related to the trapped flux by $R_c =
(2\Phi_c)^{1/2}$ in dimensionless variables. As we have shown in \S 3, all prolate cores must evolve from
parent clouds of radius $R_c \le 1.75$, but other authors chose radii larger than $2.4$.  Because
mass and flux scale differently with radius, a sufficiently large parent cloud with fixed $\alpha$ will
always be dominated by gravity, and hence evolve to an oblate configuration. Whether adopting a different
mass-to-flux distribution within the cloud will change this conclusion remains to be examined.

The fact that there exists a maximum radius, $R_{\rm max}$, for prolate cores has interesting
observational consequences.  In conventional units, this critical radius is given by
\begin{equation}
  R_{\rm max} = 0.25 pc \prn{\frac{a}{0.2km/s}}^2 \prn{\frac{P_{\rm ext}}{10^{-12} dyne/cm^2}}^{-1/2}.
\end{equation}
For typical molecular cloud cores, this value is consistent with those studied by \citet{Myers:1991}
and \citet{Ryden:1996}.  We may infer more detailed parameters of the cores if the properties of the
environment can be constrained by observations.  In particular, if we are able to obtain both the
values of $\lambda$ and
\begin{displaymath}
  \beta = 0.13 \frac{P_{\rm ext}}{10^{-12} dyne/cm^2} \prn{\frac{B_0}{10 \mu G}}^{-2},
\end{displaymath}
then we may compute the value of $\alpha$.  Inspection of Fig. \ref{crit_fig} provides an upper limit on the mass of a prolate core. As remarked earlier, the masses for prolate configurations are predicted to be lower than for oblate configurations and, hence, such cloud cores are expected to be characterized by
a lower visual extinction.

Our analysis is restricted to axisymmetric models with only poloidal fields.  These assumptions are
justified if the cores were formed via gradual contraction regulated by ambipolar diffusion \citep[however, see][for discussions on the growth of non-axisymmetric modes in sheet-like structures]{Basu:2004,Ciolek:2006}.  Relaxation
of these restrictions yields additional possibilities, which may shed light on the initial conditions and
the formation process of the molecular cores.  In order for magnetic confinement to yield prolate cores,
the parent cloud must be over dense and try to expand initially.  How to achieve such an initial state is
an intriguing question. For instance, numerical simulations suggest that collisions between turbulent
clouds would leave behind triaxial dense cores that tend to be prolate \citep{Gammie:2003, Li:2004, Offner:2009}. One
could consider the possibility that the colliding clouds created the high density region in the first
place.  After the turbulence had decayed, the subsequent expansion against an anisotropic external
magnetic pressure may facilitate the formation of the low mass prolate cores.
We reserve such a consideration for a future study.

\acknowledgments
The authors acknowledge support from the Theoretical Institute for Advanced Research in Astrophysics (TIARA) in the Academia Sinica Institute of Astronomy \& Astrophysics.  R.E.T. would like to acknowledge helpful conversations with Giles Novak.  The research of M.J.C. is supported in part by the NSC grants 95-2112-M-001-044 and 98-2112-M-001-010.

\end{document}